\documentclass{article}

\usepackage{spconf,amsmath,graphicx, xcolor}
\usepackage{amssymb, hyperref}
\usepackage{tikz, pgfplots, pgfplotstable}

\newcommand{\figref}[1]{Fig.~\ref{#1}}
\newcommand{\secref}[1]{Sec.~\ref{#1}}
\newcommand{\tabref}[1]{Table~\ref{#1}}

    \title{CONVOLUTIONAL NEURAL NETWORKS ON IRREGULAR DOMAINS BASED ON APPROXIMATE VERTEX-DOMAIN TRANSLATIONS}

    \twoauthors
    {
        B. Pasdeloup, P. Frossard
    }
    {
        Ecole Polytechnique Federale de Lausanne (EPFL) \\
        Signal Processing Laboratory (LTS4) \\
        Lausanne, Switzerland
    }
    {
        V. Gripon, J.-C. Vialatte, D. Pastor
    }
    {
        IMT Atlantique \\
        Electronics department \\
        Brest, France
    }

\begin{document}

\maketitle

\begin{abstract}
    We propose a generalization of convolutional neural networks (CNNs) to irregular domains, through the use of a translation operator on a graph structure.
    In regular settings such as images, convolutional layers are designed by translating a convolutional kernel over all pixels, thus enforcing translation equivariance.
    In the case of general graphs however, translation is not a well-defined operation, which makes shifting a convolutional kernel not straightforward.
    In this article, we introduce a methodology to allow the design of convolutional layers that are adapted to signals evolving on irregular topologies, even in the absence of a natural translation.
    Using the designed layers, we build a CNN that we train using the initial set of signals.
    Contrary to other approaches that aim at extending CNNs to irregular domains, we incorporate the classical settings of CNNs for 2D signals as a particular case of our approach.
    Designing convolutional layers in the vertex domain directly implies weight sharing, which in other approaches is generally estimated \emph{a posteriori} using heuristics.
\end{abstract}


\section{Introduction}
\label{intro}
    
\begin{figure}
        \centering
        \includegraphics[width=0.8\linewidth]{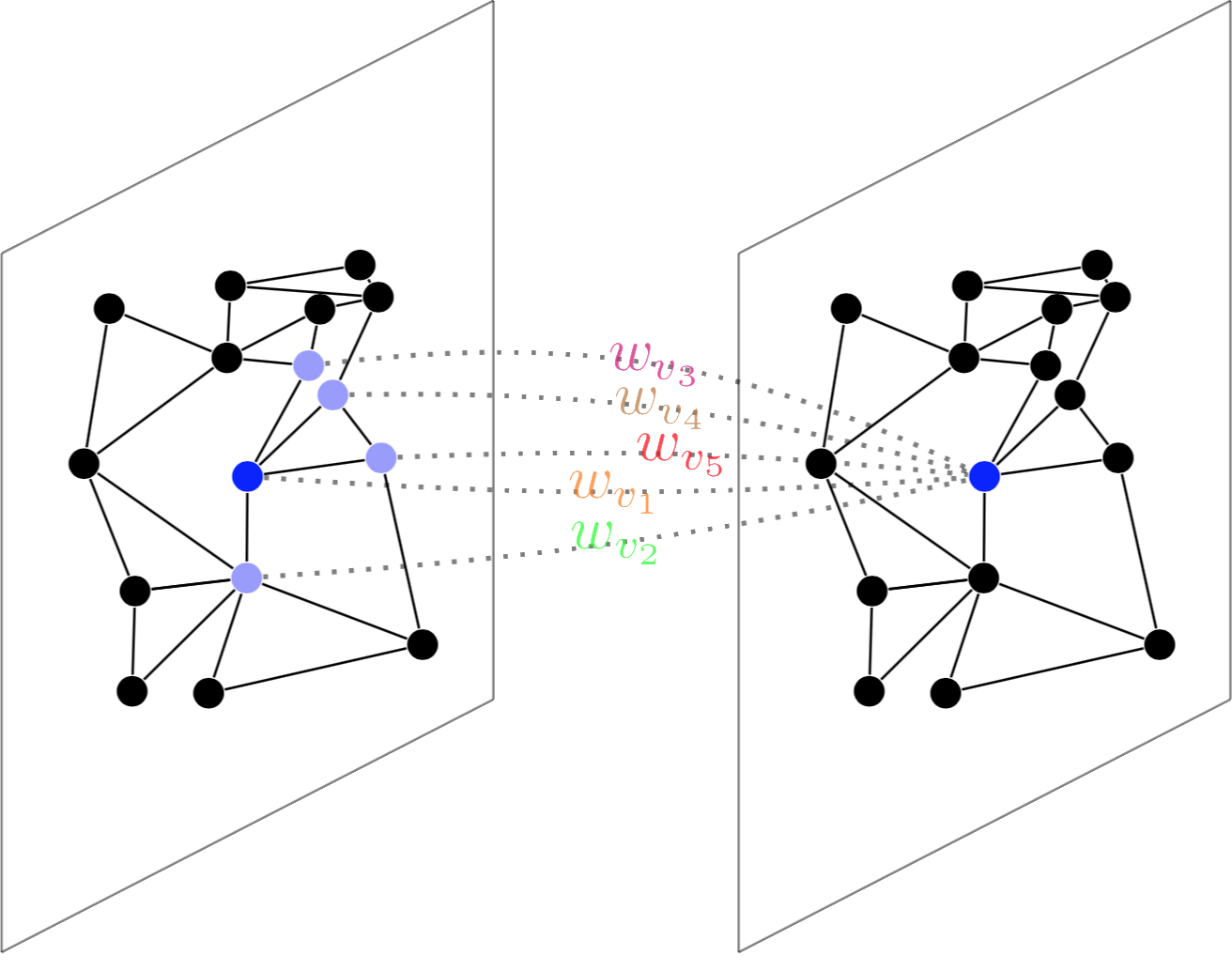}
        \caption{Example of a convolutional kernel of $5$ weights (represented with colors), localized on the dark blue variable.
        Dotted edges correspond to entries of the adjacency matrix of the layer, for which weights are given by the kernel initialization.}
        \label{exampleLayer}
    \end{figure}

    Convolutional neural networks (CNNs) have become state of the art algorithms for numerous tasks involving classification of signals.
    Their increased performance compared to dense networks is usually explained by their ability to capture local aspects of the signals under study, as well as by the fact that they are not impacted by the locations of these local elements.
    Convolutional layers are designed by initializing a small kernel of weights to train over the domain on which signals are defined, and translating it to every possible location.
    A convolutional layer can be seen as bipartite graphs of $2N$ neurons, with $N$ the dimension (or number of variables) of the signal to study (\emph{e.g.}, number of pixels in an image, or number of discrete measurements of an audio).
    Each of these variables is associated with two neurons, one input, one output.
    The adjacency matrix of the layer can be designed as follows \cite{vialatte2016generalizing}:
    For each location of the kernel, centered on a variable $l$, an edge is created in the convolutional layer between the output neuron associated with variable $l$ and each input neuron associated with a variable located under the kernel.
    Edges are then weighted according to the associated weights in the kernel, which will later be trained.
    \figref{exampleLayer} depicts a visual representation the connections resulting from one particular kernel localization.
    
    Since weights to train are set only once at kernel initialization, they are duplicated for every possible localization of the convolutional kernel.
    A small localized element in a signal defined over adjacent variables will therefore result in the same linear combination of weights whatever its localization in the signal.
    Reproducing this translation equivariance property in the context of signals defined over an irregular topology is not straightforward, as translation is not a well-defined operator on such structures.
    Identifying such a translation operator would therefore have a high impact, as it would allow reproducing the methodology introduced above, thus extending CNNs analogously to the regular space settings, with applications to analysis of signals evolving in the brain, or in a social network, among other examples.
    
    
	    Due to the potential impact of extending CNNs to irregular domains, finding such solutions has recently been a very active field of research.
	    
	    A first attempt to generalize CNNs was proposed by Bruna \emph{et al.} \cite{bruna2013spectral}, followed by Henaff \emph{et al.} \cite{henaff2015deep}, in which the authors propose to use the translation operator from the graph signal processing framework in place of Euclidean translations.
	    This operator is defined in the spectral domain associated with the graph, as a dot product between the signal spectrum and the spectrum of a Dirac signal \cite{shuman2013emerging}.
	    Due to the spectral definition, translation of a convolutional kernel on a graph using this operator does not preserve locality of the kernel.
        This problem has been addressed by Defferrard \emph{et al.} \cite{defferrard2016convolutional} by considering Chebyshev polynomial filters.
        Such filters have been used to design convolutional kernels with transformation equivariance properties of signals \cite{khasanova2017graph}.
        Levie \emph{et al.} \cite{levie2017cayleynets} introduced Cayley polynomials that also localize in the frequency domain.
        A common aspect of all these spectral-based methods is that an initial kernel is not translated at all possible locations, but re-generated on each possible variable.
        In order to enforce translation equivariance, the initial kernel weights are therefore re-affected to the localized kernel entries, using arbitrary heuristics.

        Another approach is to perform convolutions directly in the vertex domain.
        Typically, the output of these convolutions at each vertex is a function of its neighbors and a weight kernel \cite{duvenaud2015convolutional, atwood2015diffusion, monti2016geometric}.
        Locally, this operation can amount to a scalar product as in \cite{vialatte2016generalizing, niepert2016learning, puy2017unifying}, exactly like for convolutions on images.
        However, there is no explicit translation on graphs: matching vertices from these neighborhoods centered at different locations is again performed arbitrarily.
        Some strategies make use of random walks \cite{hechtlinger2017generalization}, or perform learning of kernels and matching jointly \cite{vialatte2017learning}.
        
        Due to the growing interest in graph convolutional neural networks, many additional contributions have been proposed in the recent months.
        We recommend \cite{zhanggraph} for a more complete overview.
        
    %
    
\begin{figure*}[t]
        \centering
        \begin{tikzpicture}[scale=0.6, thick]
            \node (1) [draw, circle] at (1,0) {};
            \node (2) [draw, circle] at (0,2) {};
            \node (3) [draw, circle] at (2,2) {};
            \node (4) [draw, circle] at (3,0) {};
            \node (5) [draw, circle] at (4,2) {};
            \node (6) [draw, circle] at (5,0) {};
            \node (7) [draw, circle] at (6,2) {};
            \path (1) edge (2);
            \path (1) edge (3);
            \path (2) edge (3);
            \path (1) edge (4);
            \path (3) edge (4);
            \path (4) edge (5);
            \path (5) edge (6);
            \path (5) edge (7);
            \draw [blue] (0.7,0.5) -- (0.7,1.3);
            \draw [blue] (0.9,0.5) -- (0.9,1.0);
            \draw [blue] (1.1,0.5) -- (1.1,0.8);
            \draw [blue] (1.3,0.5) -- (1.3,1.2);
            \draw [blue] (2.7,0.5) -- (2.7,1.0);
            \draw [blue] (2.9,0.5) -- (2.9,1.5);
            \draw [blue] (3.1,0.5) -- (3.1,0.8);
            \draw [blue] (3.3,0.5) -- (3.3,1.1);
            \draw [blue] (4.7,0.5) -- (4.7,0.6);
            \draw [blue] (4.9,0.5) -- (4.9,1.0);
            \draw [blue] (5.1,0.5) -- (5.1,0.7);
            \draw [blue] (5.3,0.5) -- (5.3,0.8);
            \draw [blue] (-0.3,2.5) -- (-0.3,3.0);
            \draw [blue] (-0.1,2.5) -- (-0.1,3.1);
            \draw [blue] (0.1,2.5) -- (0.1,3.4);
            \draw [blue] (0.3,2.5) -- (0.3,2.8);
            \draw [blue] (1.7,2.5) -- (1.7,2.9);
            \draw [blue] (1.9,2.5) -- (1.9,3.0);
            \draw [blue] (2.1,2.5) -- (2.1,2.9);
            \draw [blue] (2.3,2.5) -- (2.3,3.4);
            \draw [blue] (3.7,2.5) -- (3.7,2.8);
            \draw [blue] (3.9,2.5) -- (3.9,3.4);
            \draw [blue] (4.1,2.5) -- (4.1,3.5);
            \draw [blue] (4.3,2.5) -- (4.3,3.1);
            \draw [blue] (5.7,2.5) -- (5.7,2.9);
            \draw [blue] (5.9,2.5) -- (5.9,3.0);
            \draw [blue] (6.1,2.5) -- (6.1,3.1);
            \draw [blue] (6.3,2.5) -- (6.3,3.0);
            \node (translations) [draw, rectangle] at (17,-3) {\begin{tabular}{c}Translations\\identification\end{tabular}};
            \node (translationsBefore) at (6,1) {};
            \node (translationsAfter) at (17,-5) {};
            \path [dashed] (translationsBefore) edge (translations);
            \path [-latex, dashed] (translations) edge (translationsAfter);
            \node (1t1) [draw, circle] at (15,-8) {};
            \node (2t1) [draw, circle] at (14,-6) {};
            \node (3t1) [draw, circle] at (16,-6) {};
            \node (4t1) [draw, circle] at (17,-8) {};
            \node (5t1) [draw, circle] at (18,-6) {};
            \node (6t1) [draw, circle, fill=black] at (19,-8) {};
            \node (7t1) [draw, circle, fill=black] at (20,-6) {};
            \path [->, red] (1t1) edge (2t1);
            \path [dotted] (1t1) edge (3t1);
            \path [->, red] (2t1) edge (3t1);
            \path [->, red] (4t1) edge (1t1);
            \path [->, red] (3t1) edge (4t1);
            \path [dotted] (4t1) edge (5t1);
            \path [->, red] (5t1) edge (6t1);
            \path [dotted] (5t1) edge (7t1);
            \node (etc) at (17,-9) {[\dots]};
            \node (1t2) [draw, circle] at (15,-12) {};
            \node (2t2) [draw, circle, fill=black] at (14,-10) {};
            \node (3t2) [draw, circle, fill=black] at (16,-10) {};
            \node (4t2) [draw, circle] at (17,-12) {};
            \node (5t2) [draw, circle] at (18,-10) {};
            \node (6t2) [draw, circle, fill=black] at (19,-12) {};
            \node (7t2) [draw, circle, fill=black] at (20,-10) {};
            \path [dotted] (1t2) edge (2t2);
            \path [->, red] (1t2) edge (3t2);
            \path [dotted] (2t2) edge (3t2);
            \path [dotted] (4t2) edge (1t2);
            \path [dotted] (3t2) edge (4t2);
            \path [->, red] (4t2) edge (5t2);
            \path [dotted] (5t2) edge (6t2);
            \path [->, red] (5t2) edge (7t2);
            \node (1k) [draw, circle, fill=blue!40] at (6,-6) {};
            \node at (6.5, -5.5) {\textcolor{magenta}{$w_3$}};
            \node (2k) [draw, circle] at (5,-4) {};
            \node (3k) [draw, circle, fill=blue!40] at (7,-4) {};
            \node at (7.5, -3.5) {\textcolor{green}{$w_2$}};
            \node (4k) [draw, circle, fill=blue] at (8,-6) {};
            \node at (8.5, -5.5) {\textcolor{orange}{$w_1$}};
            \node (5k) [draw, circle, fill=blue!40] at (9,-4) {};
            \node at (9.5, -3.5) {\textcolor{brown}{$w_4$}};
            \node (6k) [draw, circle] at (10,-6) {};
            \node (7k) [draw, circle] at (11,-4) {};
            \path (1k) edge (2k);
            \path (1k) edge (3k);
            \path (2k) edge (3k);
            \path (1k) edge (4k);
            \path (3k) edge (4k);
            \path (4k) edge (5k);
            \path (5k) edge (6k);
            \path (5k) edge (7k);
            \node (cnnLayer) [draw, rectangle] at (8,-9) {\begin{tabular}{c}Kernel\\translation\end{tabular}};
            \node (cnnLayerBefore) at (13,-9) {$\left\{\begin{array}{l}~\\~\\~\\~\\~\\~\\~\\~\\~\\~\end{array}\right.$};
            \node (cnnLayerAfter) at (4,-9) {};
            \node (cnnLayerTop) at (8,-7) {};
            \path [dashed] (cnnLayerBefore) edge (cnnLayer);
            \path [dashed] (cnnLayerTop) edge (cnnLayer);
            \path [-latex, dashed] (cnnLayer) edge (cnnLayerAfter);
            \node (cnnL1) [draw, circle] at (1, -6) {};
            \node (cnnL2) [draw, circle] at (1, -7) {};
            \node (cnnL3) [draw, circle] at (1, -8) {};
            \node (cnnL4) [draw, circle] at (1, -9) {};
            \node (cnnL5) [draw, circle] at (1, -10) {};
            \node (cnnL6) [draw, circle] at (1, -11) {};
            \node (cnnL7) [draw, circle] at (1, -12) {};
            \node (cnnR1) [draw, circle] at (3, -6) {};
            \node (cnnR2) [draw, circle] at (3, -7) {};
            \node (cnnR3) [draw, circle] at (3, -8) {};
            \node (cnnR4) [draw, circle] at (3, -9) {};
            \node (cnnR5) [draw, circle] at (3, -10) {};
            \node (cnnR6) [draw, circle] at (3, -11) {};
            \node (cnnR7) [draw, circle] at (3, -12) {};
            \path [orange] (cnnL1) edge (cnnR1);
            \path [brown] (cnnL4) edge (cnnR1);
            \path [magenta] (cnnL1) edge (cnnR2);
            \path [orange] (cnnL2) edge (cnnR2);
            \path [brown] (cnnL4) edge (cnnR2);
            \path [green] (cnnL1) edge (cnnR3);
            \path [brown] (cnnL4) edge (cnnR3);
            \path [orange] (cnnL3) edge (cnnR3);
            \path [magenta] (cnnL2) edge (cnnR4);
            \path [green] (cnnL3) edge (cnnR4);
            \path [orange] (cnnL4) edge (cnnR4);
            \path [brown] (cnnL5) edge (cnnR4);
            \path [green] (cnnL4) edge (cnnR5);
            \path [orange] (cnnL5) edge (cnnR5);
            \path [brown] (cnnL6) edge (cnnR5);
            \path [green] (cnnL5) edge (cnnR6);
            \path [orange] (cnnL6) edge (cnnR6);
            \path [green] (cnnL5) edge (cnnR7);
            \path [orange] (cnnL7) edge (cnnR7);
            \node (cnnCreation) [draw, rectangle] at (-3,-9) {\begin{tabular}{c}Assemble\\layers\end{tabular}};
            \node (cnnCreationBefore) at (0,-9) {};
            \node [inner sep=8] (cnnCreationAfter) at (-3,-6) {CNN};
            \path [dashed] (cnnCreationBefore) edge (cnnCreation);
            \path [-latex, dashed] (cnnCreation) edge (cnnCreationAfter);
            \node (cnnTraining) [draw, rectangle] at (-3,-3) {Training};
            \node (cnnTrainingTop) at (0,-1) {};
            \node (cnnTrainingAfter) at (-7,-3) {\begin{tabular}{c}Trained\\CNN\end{tabular}};
            \path [dashed] (cnnTrainingTop) edge (cnnTraining);
            \path [dashed] (cnnCreationAfter) edge (cnnTraining);
            \path [-latex, dashed] (cnnTraining) edge (cnnTrainingAfter);
            \node at (11.5,-0.5) {\textbf{(a)}};
            \node at (9,-6.5) {\textbf{(b)}};
            \node at (8,-10.5) {\textbf{(c)}};
            \node at (-3,-10.5) {\textbf{(d)}};
            \node at (-1,-3) {\textbf{(e)}};
        \end{tikzpicture}
        \caption[]
        {
            Introduced methodology to create a CNN adapted to given signals.
            \textbf{(a)} We identify some translation operators on the graph on which signals are defined;
            \textbf{(b)} We initialize a convolutional kernel on the graph, defined as a subset of close vertices associated with weights $(w_i)_i$ to train.
                         Here, we choose to consider the most central vertex (in blue) as the kernel center, and to include its direct neighborhood (in mauve) in the kernel vertices;
            \textbf{(c)} Using the previously found translations, we translate the convolutional kernel so that it gets centered on every possible vertex.
                         Once the kernel has been localized on every vertex, we can define a corresponding convolutional layer in the manner of \cite{vialatte2016generalizing}.
                         Mapping of the various weights, represented by the colors in the resulting layer, is given directly by the translation of the initial kernel;
            \textbf{(d)} By considering various initializations of kernels, we obtain different convolutional layer models.
                         A CNN can then be constructed by assembling them to form a complex network;
            \textbf{(e)} Finally, the CNN is trained using signals on the graph.
                         The trained network can then be used in order to classify new signals.
        }
        \label{methodologySummary}
    \end{figure*}
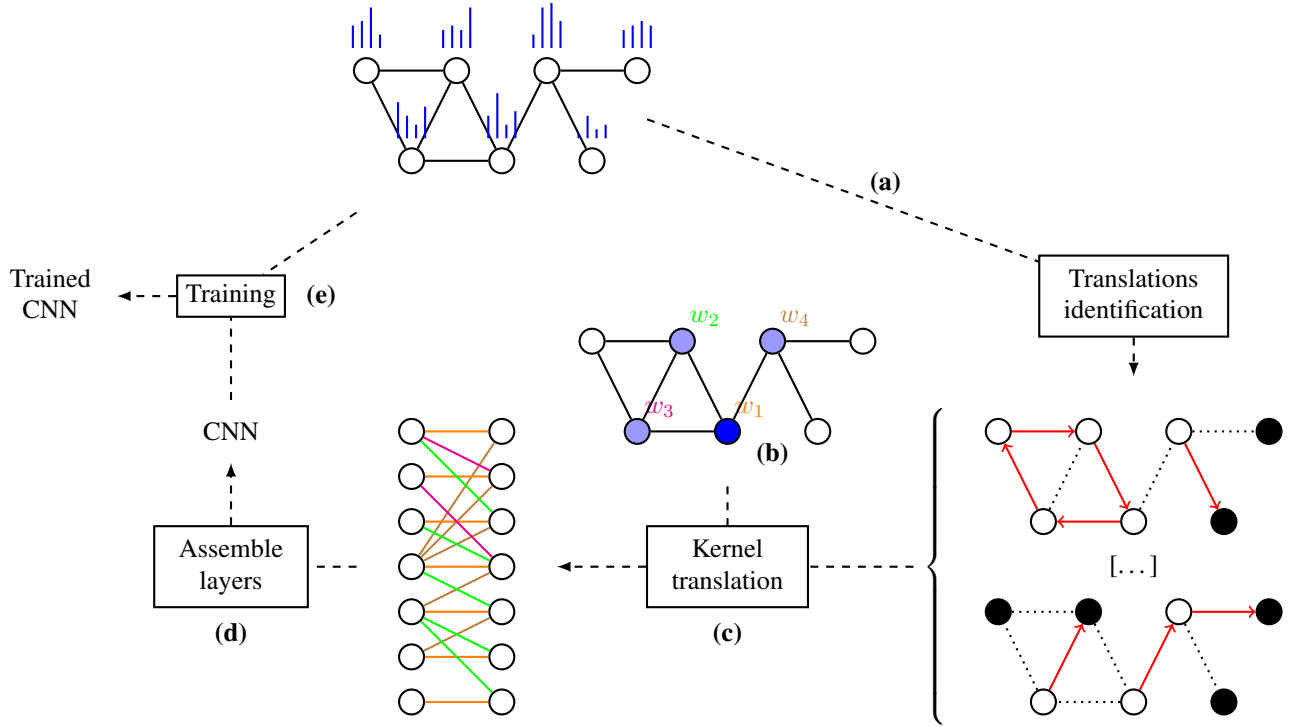

        
        Contrary to existing approaches that aim at finding a proper affectation of weights to kernels generated over each variable, we introduce in this article a novel methodology to mimic classical CNNs design, through the use of translations in the vertex domain.
        We then instanciate a single convolutional kernel on the graph, with associated weights to train, and translate it to every possible vertex.
        This methodology has the advantage to directly associate weights with the edges in a convolutional layer, without the need for heuristics \cite{vialatte2016generalizing}.
        Also, convolutional layers on 2D signals can be seen as a particular case of our approach, in which the underlying graph is a grid.
        A visual representation of the proposed methodology is given in \figref{methodologySummary}.

    The remaining of this article is organized as follows:
    In \secref{methodo}, we give details on the introduced methodology.
    Then, in \secref{experiments}, we design a convolutional layer from a set of functional magnetic resonance imaging (fMRI) signals.
    Finally, \secref{conclu} concludes.
    
\section{METHODOLOGY}
\label{methodo}

    As introduced in \secref{intro}, we are interested in identifying a translation operator on a graph, that models the domain on which signals evolve.
    In some cases, such a graph is not provided, as there is no known ground truth.
    Examples include brain signals, seismic sensors, or weather stations, for which variables are known but no connectivity is directly implied.
    For such cases, graph inference methods have been developed (\emph{e.g.}, \cite{kalofolias2016learn, pasdeloup2017characterization}) that allow defining a topology that is adapted to signals.
    In this work, we assume the graph to be either given or inferred using such approaches.
    
    In \cite{grelier2016neighborhood, pasdeloup2017translations}, authors have proposed a definition for translations that are analogous to Euclidean translations on the grid graphs, while not making use of the metric space.
    Let $\mathcal{G} = \langle \mathcal{V}, \mathcal{E} \rangle$ be a graph.
    A translation as introduced in \cite{pasdeloup2017translations} is a function $\psi : \mathcal{V} \mapsto \mathcal{V} \cup \{\bot\}$, where $\bot$ allows the disappearance of signal entries, such that:
        \begin{itemize}
            \item $\psi$ is injective for each vertex whose image is in $\mathcal{V}$, \emph{i.e.}, $\forall v_1, v_2 \in \mathcal{V} : (\psi(v_1) = \psi(v_2) \neq \bot) \Rightarrow (v_1 = v_2)$;
            \item $\psi$ is edge-constrained (EC), \emph{i.e.}, $\forall v \in \mathcal{V} : \big((v, \psi(v)) \in \mathcal{E}\big) \vee \big(\psi(v) = \bot\big)$;
            \item $\psi$ is strongly neighborhood-preserving (SNP), \emph{i.e.}, $\forall v_1, v_2 \in \mathcal{V} : \big((v_1, v_2) \in \mathcal{E} \Leftrightarrow (\psi(v_1), \psi(v_2)) \in \mathcal{E}\big) \vee (\psi(v_1) = \bot) \vee (\psi(v_2) = \bot)$.
        \end{itemize}
        
        Such operators are very well adapted to the problem of designing convolutional layers for graphs.
        As a matter of fact, these functions being injective directly implies the mapping of weights in the design of a layer.
        Additionally, the SNP property enforces the convolutional kernel to remain similar as it gets translated over the graph, which enforces the translation equivariance property we want to achieve.
        
        The authors have however shown that identification of these translations is an NP-complete problem.
        To tackle this problem, they have proposed a notion of approximate translations, that allow controlled violations of the properties listed previously.
        
        In this article, we make use of these approximate translations to allow translating a single convolutional kernel to all possible vertices of the graph.
        We proceed as follows:
        \begin{enumerate}
            \item We initialize a kernel $\{v_i:w_i\}_i, v_i \in \mathcal{V}$ --- defined as an association of weights $w_i$ to train to a selected subset of vertices --- on one of the most central vertices $v_1$ in the graph, considering the proximity centrality \cite{sabidussi1966centrality};
            \item For each neighbor $v_2$ of $v_1$, we identify an approximate translation $\psi_{v_1 \rightarrow v_2}$ allowing to center the kernel on $v_2$;
            \item For every possible new location of the kernel, we start again from 2. until we reach a fixed point of the process.
                  When multiple potential kernels are centered on the same vertex $v_3 \in \mathcal{V}$, we only keep the one obtained by an approximate translation $\psi_{v_1 \rightarrow v_3}$ that minimizes the score $s(\psi_{v_1 \rightarrow v_3})$ defined in \cite{pasdeloup2017translations} (5), which measures the total deformation of the signal along the path from $v_1$ to $v_3$.
        \end{enumerate}
        
        The algorithm stops when the best kernel (according to the score function) has been found for every possible center.
        As the number of elementary paths between any pair of vertices in a connected graph is limited, the algorithm terminates.
        Finally, once the kernel has been localized on every possible vertex with minimum deformation, we create the adjacency matrix of the convolutional layer as in the classical settings of CNNs over regular spaces.

\section{EXPERIMENTS}
\label{experiments}
        
        \begin{table*}
            \centering
            \def\arraystretch{1.1}
            \begin{tabular}{c|c|c|c|c|c|c|c|c|}
                \cline{2-9}
                & $\mathcal{G}_{geo}$ & Chebnet \cite{defferrard2016convolutional} & MLP & SVM & LR ($C=1$, $l_1$) & LR ($C=50$, $l_1$) & (LR $C=1$, $l_2$) & Ridge \\
                \hline
                \multicolumn{1}{|c|}{Subject 1} & $56\%$ & $\boldsymbol{60\%}$ & $54\%$ & $50\%$ & $51\%$ & $47\%$ & $49\%$ & $35\%$ \\
                \hline
                \multicolumn{1}{|c|}{Subject 2} & $52\%$ & $\boldsymbol{56\%}$ & $44\%$ & $42\%$ & $48\%$ & $47\%$ & $47\%$ & $34\%$ \\
                \hline
                \multicolumn{1}{|c|}{Subject 3} & $47\%$ & $\boldsymbol{51\%}$ & $48\%$ & $35\%$ & $44\%$ & $38\%$ & $41\%$ & $28\%$ \\
                \hline
                \multicolumn{1}{|c|}{Subject 4} & $53\%$ & $\boldsymbol{54\%}$ & $42\%$ & $32\%$ & $34\%$ & $28\%$ & $34\%$ & $22\%$ \\
                \hline
                \multicolumn{1}{|c|}{Subject 5} & $42\%$ & $\boldsymbol{45\%}$ & $39\%$ & $31\%$ & $42\%$ & $34\%$ & $31\%$ & $23\%$ \\
                \hline
                \multicolumn{1}{|c|}{Subject 6} & $49\%$ & $\boldsymbol{51\%}$ & $49\%$ & $40\%$ & $42\%$ & $40\%$ & $38\%$ & $31\%$ \\
                \hline
                \hline
                \multicolumn{1}{|c|}{Average} & $49.9\%$ & $\boldsymbol{52.8\%}$ & $45.9\%$ & $38.2\%$ & $43.6\%$ & $38.9\%$ & $39.7\%$ & $28.8\%$ \\
                \hline
            \end{tabular}
            \caption
            {
                Results of our method with the graph $\mathcal{G}_{geo}$ on the Haxby dataset.
                Other results are obtained with the Chebnet CNN using Chebyshev polynomials \cite{defferrard2016convolutional}, a multilayer perceptron (MLP), and with the classifiers adapted from \cite{nilearn2017haxby}.
            }
            \label{haxbyResults}
        \end{table*}
        
        In order to assess quality of the proposed method, we have chosen to consider a dataset of brain signals, acquired through functional magnetic resonance imaging (fMRI), on which classical CNNs cannot be used.
        
        The Haxby dataset of fMRI signals \cite{haxby2001distributed} consists of measurements of the brain activity during a visualization task, in which images of $8$ classes are shown to $6$ subjects during $12$ sessions.
        For each subject, activity in $163,840$ voxels is measured.
        A parcellation is then applied to group the vertices using functional or geographical properties, in order to reduce the number of variables to $444$ \cite{bellec2010multi}.
        As these locations are associated with coordinates in the brain, we choose here to consider a graph $\mathcal{G}_{geo}$ by computing the Euclidean distances between vertices, and keeping the $6$ highest entries per row, then symmetrizing.
        
        In order to have a training and a test set of signals, we have splitted the $12$ sessions into a training set of $10$ sessions and a test set of $2$ sessions.
        To provide a baseline to compare with, we have adapted the classifiers in \cite{nilearn2017haxby} in order to have them perform classification of $8$ classes at once.
        Additionally, we compare with a multilayer perceptron, made of three layers with parameters found with grid search.
        Finally, we also provide the results of Chebnet, a CNN with convolutional layer designed using Chebyshev polynomials \cite{defferrard2016convolutional} of order $5$.
        
        The CNN we consider consists of three layers with ReLU non-linearities: one convolutional layer and two dense layers, followed by some dropout, with parameters found by grid search.
        This design has been chosen to be close to the MLP, in order to evaluate a gain in performance related to the use of convolutional layers.

        Results obtained for all these classifiers are depicted in \tabref{haxbyResults}.
        They demonstrate that the introduced method provides improved results compared to a MLP or a simple classifier.
        However, Chebnet offers better performance in comparison to ours.
        While the number of weights to train is similar between Chebnet and our approach, we explain that difference by the number of vertices from which signal information is drawn.
        A Chebyshev polynomial of order $5$ draws information from a larger set of vertices --- those that are at most $5$ hops away from the kernel center --- than our method does, which appears to have more impact on classification results than finding a proper mapping.

        It is still worth noting that the gap in performance is not significantly high, which encourages us continue developing the method.
        We believe it to be able to outperform other classifiers especially on datasets for which orientation of features within signals makes sense, like when considering an irregular sampling of a manifold.

\section{CONCLUSIONS}
\label{conclu}

    In this article, we have introduced a methodology to extend CNNs to signals evolving on irregular domains.
    Contrary to other works in the domain, our approach aims at finding translations on the graph in order to reproduce the design methodology of classical CNNs over regular spaces.
    The translations used are injective function that preserve the shape of the convolutional kernel, thus implicitely setting the mapping of the weights and enforcing translation equivariance.
    It therefore includes the design of classical convolutional layers as a particular case, in which the underlying graph is a grid.
    Experiments have shown applicability of the proposed methodology on signals evolving in the human brain, for which we outperform simple classifiers, while being comparable with Chebyshev polynomials-based approaches.

    Recent results have been developed based on a previous version of the present article \cite{lassance2018matching}, in which the authors propose a pooling algorithm to complement the introduced convolutional layers, as well as stride convolutions and data augmentation techniques.
    Improved CNNs using these additional tools have allowed to outperform the performance of Chebnet on another dataset of fMRI signals, which encourages us to continue developing the introduced approach.
    One possible directions for doing so is to study the impact of the convolutional kernel shape on performance, and design less arbitrary shapes, for instance using dictionary learning approaches.

\bibliographystyle{IEEEbib}
\bibliography{refs}

\end{document}